\documentclass[twocolumn, 10pt]{IEEEtran}
\usepackage{graphicx}
\usepackage{amssymb}
\usepackage{mathtools}
\usepackage{dsfont}
\usepackage{cite}
\usepackage{stfloats}
\usepackage{subfigure}
\usepackage{psfrag}
\usepackage[mathscr]{euscript}
\usepackage{acronym}  
\usepackage{amsmath}
\usepackage{algorithm}
\usepackage[noend]{algpseudocode}
\usepackage{booktabs}
\usepackage{bbm}

\makeatletter
\def\BState{\State\hskip-\ALG@thistlm}
\makeatother

\usepackage{float}
\usepackage{balance}

\acrodef{CCDF}{complementary cumulative distribution function}
\acrodef{CF}{characteristic function}
\acrodef{PPP}{Poisson point processe}
\acrodef{RV}{random variable}
\acrodef{i.i.d.}{independent and identically distributed}
\acrodef{PDF}{probability distribution function}
\acrodef{CDF}{cumulative distribution function}
\acrodef{ch.f.}{characteristic function}
\acrodef{AWGN}{additive white Gaussian noise}
\acrodef{SNR}{signal-to-noise ratio}
\acrodef{LRT}{likelihood ratio test}
\acrodef{DRT}{distance ratio test}
\acrodef{GLRT}{generalized likelihood ratio test}
\acrodef{CRLB}{Cram\'{e}r-Rao lower bound}
\acrodef{CRB}{Cram\'{e}r-Rao bound}
\acrodef{ZZLB}{Ziv-Zakai lower bound}
\acrodef{ZZB}{Ziv-Zakai bound}
\acrodef{LOS}{line-of-sight}
\acrodef{ToF}{time-of-flight}
\acrodef{NLOS}{non-line-of-sight}
\acrodef{GDOP}{geometric dilution of precision}
\acrodef{GPS}{Global Positioning System}
\acrodef{FIM}{Fisher information matrix}
\acrodef{PEB}{position error bound}
\acrodef{SPEB}{squared position error bound}
\acrodef{TOA}{time-of-arrival}
\acrodef{TOF}{time-of-flight}
\acrodef{WSN}{wireless sensor network}
\acrodef{MAC}{medium access control}
\acrodef{RSS}{received signal strength}
\acrodef{WAF}{wall attenuation factor}
\acrodef{TDOA}{time difference-of-arrival}
\acrodef{RF}{radiofrequency}
\acrodef{RTT}{round-trip time}
\acrodef{AOA}{angle-of-arrival}
\acrodef{MF}{matched filter}
\acrodef{ED}{energy detector}
\acrodef{ML}{maximum likelihood}
\acrodef{MSE}{mean-square error}
\acrodef{RMSE}{root-mean-square error}
\acrodef{LEO}{localization error outage}
\acrodef{ppm}{part-per-million}
\acrodef{ACK}{acknowledge}
\acrodef{UWB}{Ultrawide bandwidth}
\acrodef{TNR}{threshold-to-noise ratio}
\acrodef{LS}{least squares}
\acrodef{IR-UWB}{impulse radio UWB}
\acrodef{FCC}{Federal Communications Commission}
\acrodef{TH}{time-hopping}
\acrodef{PPM}{pulse position modulation}
\acrodef{MUI}{multi-user interference}
\acrodef{PDP}{power delay profile}
\acrodef{BPZF}{band-pass zonal filter}
\acrodef{SIR}{signal-to-interference ratio}
\acrodef{SINR}{signal-to-interference-plus-noise ratio}
\acrodef{RFID}{radio frequency identification}
\acrodef{WPAN}{wireless personal area network}
\acrodef{WWB}{Weiss-Weinstein bound}
\acrodef{DP}{direct path}
\acrodef{MF}{matched filter}
\acrodef{MMSE}{minimum-mean-square-error}
\acrodef{SBS}{serial backward search}
\acrodef{SBSMC}{serial backward search for multiple clusters}
\acrodef{NBI}{narrowband interference}
\acrodef{WBI}{wideband interference}
\acrodef{INR}{interference-to-noise ratio}
\acrodef{CR}{channel response}
\acrodef{CIR}{channel impulse response}
\acrodef{CR}{channel  response}
\acrodef{RADAR}{radar}
\acrodef{MUR}{Multistatic radar}
\acrodef{JBSF}{jump back and search forward}
\acrodef{HDSA}{high-definition situation-aware}
\acrodef{RRC}{root raised cosine}
\acrodef{ST}{simple thresholding}
\acrodef{BTB}{Bellini-Tartara bound}
\acrodef{P-Max}{$P$-Max}  
\acrodef{MIMO}{multiple-input multiple-output}
\acrodef{MAP}{maximum a posteriori}
\acrodef{FG}{factor graph}
\acrodef{OP}{outage probability}
\acrodef{WED}{wall extra delay}
\acrodef{RMS}{root mean square}
\acrodef{SPAWN}{sum-product algorithm over a wireless network}
\acrodef{MDD}{minimum distance distribution}
\acrodef{MAP}{maximum a posteriori probability}
\acrodef{SAP}{small cell access point}
\acrodef{UE}{user equipment}
\acrodef{MBS}{macro cell base station}
\acrodef{UER}{\ac{UE} Relay}
\acrodef{D2D}{device-to-device}
\acrodef{MBS}{macro base station}
\acrodef{CSI}{channel state information}
\acrodef{OGR}{outage guard region}
\acrodef{FUR}{feasible UER region}
\acrodef{EHR}{energy harvesting region}
\acrodef{EH}{energy harvesting}
\acrodef{D2D-EHSN}{D2D communication provided \ac{EH} small cell network}
\acrodef{D2D-EHHN}{D2D communication provided \ac{EH} heterogeneous network}
\acrodef{3GPP}{3rd Generation Partnership Project}
\acrodef{BS}{base station}
\acrodef{DF}{decode and forward}
\acrodef{CCDF}{complementary cumulative distribution function}
\acrodef{ZF}{zero forcing}
\acrodef{RZF}{regularized zero forcing}
\acrodef{WLLN}{weak law of large number}
\acrodef{SLLN}{strong law of large numbers}
\acrodef{TDD}{Time-division duplex}
\acrodef{EE}{energy efficiency} 
\acrodef{HetNet}{heterogeneous network} 
\acrodef{SCP}{Single Cell Processing}
\acrodef{CBF}{Coordinated Beamforming}
\usepackage{color}
\usepackage{dsfont}
\usepackage{bbm}

\DeclareMathAlphabet{\mathsf}{OML}{cmbr}{m}{it}

\newtheorem{theorem}{\bf Theorem}
\newtheorem{lemma}{\bf Lemma}
\newtheorem{corollary}{\bf Corollary}

\newcommand{\bd}{\begin{description}}
\newcommand{\ed}{\end{description}}
\newcommand{\be}{\begin{enumerate}}
\newcommand{\ee}{\end{enumerate}}
\newcommand{\bi}{\begin{itemize}}
\newcommand{\ei}{\end{itemize}}
\newcommand{\bl}{\begin{list}}
\newcommand{\el}{\end{list}}
\newcommand{\bt}{\begin{tabbing}}
\newcommand{\et}{\end{tabbing}}

\newcommand{\paperTitle}{Analysis of the Age of Information in Age-Threshold Slotted ALOHA }

\begin{document}

{
\title{\paperTitle}

\author{
  \vspace{0.2cm}
      Howard~H.~Yang$^\dagger$, 
      Nikolaos Pappas$^\star$,
      Tony Q. S. Quek$^\ddagger$, 
      and Martin Haenggi$^\mathsection$ \\
       
    $^\dagger$ \textit{ZJU-UIUC Institute, Zhejiang University, Haining 314400, China }\\
    $^\star$ \textit{Department of Computer and Information Science, Linköping University, Linköping 58183, Sweden}\\
    $^\ddagger$ \textit{ISTD Pillar, Singapore University of Technology and Design, Singapore 487372, Singapore} \\
    $^\mathsection$ \textit{Department of Electrical Engineering, University of Notre Dame, Notre Dame, IN 46556, USA} \\   
    \vspace{0.13cm}
    (Invited Paper)\\
    \vspace{-0.75cm}

}

\maketitle
\acresetall
\thispagestyle{empty}
\begin{abstract}
We investigate the performance of a random access network consisting of source-destination dipoles. 
The source nodes transmit information packets to their destinations over a shared spectrum. 
All the transmitters in this network adhere to an age threshold slotted ALOHA (TSA) protocol: every source node remains silent until the age of information (AoI) reaches a threshold, after which the source accesses the radio channel with a certain probability. 
We derive a tight approximation for the signal-to-interference-plus-noise ratio (SINR) meta distribution and verify its accuracy through simulations. 
We also obtain analytical expressions for the average AoI. Our analysis reveals that when the network is densely deployed, employing TSA significantly decreases the average AoI. The update rate and age threshold must be jointly optimized to fully exploit the potential of the TSA protocol. 
\end{abstract}

\acresetall
\section{Introduction}\label{sec:intro}
The next generation (6G) wireless system \cite{TarKhaWon:20WCMAG,DanAmiShi:20NE} is envisioned to not just provide fast communications, but also enable real-time sensing and even decision making capabilities, necessitating the deployment of massively distributed, ad hoc networks in which sensors collect and send information updates to data fusion centers. 
In light of the timeliness requirement, the transmissions shall be conducted with minimal coordination overhead using simple random access schemes.

To quantify timeliness, the notion of age of information (AoI) has been proposed in \cite{KauYatGru:12,Pappas2023age}, which is evaluated from the receiver's perspective, measuring the time elapsed since the latest information packet received at the destination has been generated from the source.
It is proven in \cite{TalMod:21IT} that optimizing the AoI metric differs intrinsically from classical networking formulations that maximize throughput or minimize delay. 
Consequently, this has triggered new research focused specifically on AoI optimization.
The concept of AoI, on the other hand, also facilitates networking designs \cite{AbdPapDhi:19CMAG}, where age-relevant information can be leveraged to construct distributed transmission policies to significantly improve the performance of a wireless system. 
A particular example is the age threshold slotted ALOHA (TSA) protocol \cite{YavUys:21} which modifies the slotted ALOHA by incorporating an age threshold into it.
Under this protocol, every source node remains silent until the AoI reaches a predefined threshold, after which the source accesses the radio channel with a constant probability, as in the conventional slotted ALOHA. 

The present paper studies the effect of the TSA protocol on various network performance metrics through the lens of the signal-to-interference-plus-noise ratio (SINR) model, under which the packet transmission over a wireless link is successful only if the SINR at the receiver surpasses a decoding threshold. 
The SINR model is close to reality but complex for analysis, as the transmissions of a single node impact all other nodes, resulting in a space-time coupling of the involved nodes. 
As such, the traffic generation pattern of a source has a composite influence on the AoI by affecting the status updating interval and the time spent in packet delivery.
This effect is linked through the interference the nodes cause. 

\subsection{Contributions}
The main contributions of this paper are summarized below. 
\begin{itemize}
  \item We establish a mathematical framework for investigating the effects of TSA protocol in large-scale wireless networks. Our model encompasses key features such as the status updating rate, age threshold, channel fading, deployment density, and co-channel interference.
  \item We derive a tight approximation for the SINR meta distribution \cite{Hae:21CL-1,Hae:21CL-2} and verify its accuracy through simulations. The result is given in the form of a fixed-point functional equation, accounting for the spatial-temporal interactions amongst the transmitters. 
  \item We also obtain an analytical expression for the time average AoI. We provide a set of special case studies to explore the effect of TSA on the network age performance and garner useful design insights. 
  \item Our analysis demonstrates the advantages of TSA over the conventional slotted ALOHA. Specifically, when the deployment density increases, the AoI under slotted ALOHA surges sharply. In contrast, TSA substantially reduces the average AoI, endowing the network with timeliness. Although one needs to adequately optimize the update rate and age threshold to reap the full harvest.  
\end{itemize}
\subsection{Related Works}
Here, we briefly review the AoI related studies in wireless networks based on the SINR model \cite{HuZhoZha:18,ManAbdDhi:20,EmaElSBau:20,YanXuWan:19,YanAraQue:20TWC,ManCheAbd:20b,YanSweQue:21,YanArafaQue:19,SonYanSha:21TMC,YueYanZha:22JSAC}. 
These works combine stochastic geometry and queueing theory to develop theoretical models that characterize the effects from spatial and temporal attributes, facilitating the AoI analysis in a random access network.
Particularly, \cite{HuZhoZha:18} uses the favorable/dominant argument to derive upper and lower bounds of the AoI, and the upper bound is improved by \cite{ManAbdDhi:20} through a careful reconstruction of the dominant system.
By integrating a discrete-time Markov chain with stochastic geometry, \cite{EmaElSBau:20} derives the time average of peak AoI in large-scale wireless networks, while \cite{YanXuWan:19} and \cite{YanAraQue:20TWC} investigate the network average AoI under different buffer configurations (unit size versus infinite capacity) and queueing disciplines (first-come first-serve and last-come first-serve with preemption or replacement).
Additionally, \cite{YueYanZha:22JSAC} examines the AoI performance of random access networks operating under frame slotted ALOHA-based protocols, \cite{YueYanZha:22Globecom} extends the performance metric from linear AoI to that under a non-linear cost function, and \cite{ManCheAbd:20b} presents a comprehensive study on the interplay between throughput and AoI in a cellular-based IoT network.
Several distributed algorithms have been proposed in \cite{YanSweQue:21,YanArafaQue:19,SonYanSha:21TMC}, capitalizing on the source nodes' local observations to devise status update rate \cite{YanSweQue:21}, channel access probability \cite{YanArafaQue:19}, and/or power control scheme \cite{SonYanSha:21TMC} that adapt in accordance with the transmitters' local communication environment and optimize AoI. 
Nevertheless, these existing results primarily pertain to the conventional slotted ALOHA protocol, whilst the effects of the age threshold remain unexplored. 

\section{System Model}\label{sec:sysmod}
\subsection{Spatial Configuration}
We consider a wireless network containing a set of source nodes and their intended destinations located in the Euclidean plane.
The source nodes are scattered according to a homogeneous PPP $\tilde{\Phi}$ of spatial density $\lambda$. 
A source located at $X_i \in \tilde{\Phi}$, $i \geq 1$, has a dedicated receiver at $y_i$ that is at a distance $r$ from it and oriented in a random direction. 
Then, according to the displacement theorem \cite{BacBla:09}, the locations of the receivers, denoted as $\bar{\Phi} = \{y_i\}_{i=1}^\infty$, also constitute a homogeneous PPP with the same density.
Every source tries to communicate its latest status to the corresponding destination. 
The status information of each source is encapsulated into information packets and transmitted over a shared spectrum.
When a source node sends out information packets, it transmits at a fixed power. 
We assume that the channel between any pair of nodes is affected by the Rayleigh fading, which varies independently across time slots, and path loss that follows the power law.
We also assume the received signal is subjected to white Gaussian thermal noise. 

\subsection{Temporal Pattern}
We partition the time into equal-length intervals, each being the duration to transmit a single packet.
We assume the network is synchronized.
We consider each source node employs the \textit{generate-at-will} model \cite{TalKarMod:18} for the status update. 
Particularly, if a node decides to transmit, it generates a new sample at the beginning of the time slot and sends the information packet to the destination immediately. 
By the end of the same time slot, the packet is successfully decoded if the received SINR exceeds a decoding threshold; otherwise, the transmission fails. 
The delivery of packets, therefore, incurs a delay of one time slot. 
Since the time scale of fading and packet transmission is much smaller than that of the spatial dynamics, we assume the network topology is static, i.e., an arbitrary point pattern is realized at the beginning and remains unchanged over the time domain.
Additionally, we assume the time starts at $t = - \infty$, hence the system dynamic has reached the steady state at $t=0$. 

\subsection{Age of Information}
We put the main focus upon the notion of AoI in this paper. 
AoI grows linearly with time in the absence of new updates at the destination, and it reduces to the time elapsed since the generation of the delivered packet when a new information packet is received. 
We add a receiver located at the origin $o$ to the point process $\bar{\Phi}$. 
We also add its tagged transmitter, denoted by $X_0$, to $\tilde{\Phi}$.
Then, by applying Slivnyak's theorem \cite{BacBla:09}, it is sufficient to concentrate our analysis on this transmitter-receiver pair. 
We slightly abuse the definition and coin this receiver as the \textit{typical} one even before averaging over the point processes.{\footnote{A detailed discussion on the concept of typicality in stochastic spatial models is available in \cite{Hae:20Blog}.}}
We also refer to the wireless connection between the typical receiver and its tagged transmitter as the \textit{typical} link. 
The age evolution process over the typical link can be written as 
\begin{align}
\Delta_0( t ) =  t - G_0(t),
\end{align}
where $G_0(t)$ is the generation time of the latest packet delivered over this link at time $t$.
A pictorial example is provided in Fig.~\ref{fig:AoIMod_V1}.

\subsection{Transmission Protocol}
In this work, we employ the slotted ALOHA protocol in conjunction with an age threshold to control the channel access of the source-destination pairs. 
Specifically, every source node stays silent until its AoI reaches a threshold, denoted by $A$, upon which the source node turns on the status updating mode: it generates a fresh sample with probability $\eta$ at the beginning of each time slot and immediately sends that packet to the destination. 
If the transmission succeeds, the receiver feeds back an ACK to the source, and the age is reset to one. Otherwise, the source will generate a new packet in the next time slot, again with probability $\eta$. 
Note that there are no retransmissions of the undelivered packets in this mechanism. 

\subsection{Performance Metric}
We adopt the average AoI over the network as our primary performance metric.
Concretely, we denote by $\Delta_j(t)$ the AoI of link $j$ at time slot $t$, and define the time-average AoI over this link as 
\begin{align}
\bar{\Delta}_j = \limsup_{ T \rightarrow \infty } \frac{1}{ T } \sum_{ t=1 }^T  \Delta_j( t ).
\end{align}
Then, the network average AoI is given by 
\begin{align}
\bar{\Delta} =  \limsup_{R \rightarrow \infty } \frac{ \sum_{j: X_j \in B(o, R)  } \bar{ \Delta }_j  }{ \lambda \pi R^2 }
\end{align}
where $B(o,R)$ represents a disk centered at the origin with radius $R$.

Since the point process of source-destination pairs is stationary, the AoI averaged over different links across a static realization of the network is equivalent to taking the expectation of the time-average AoI at the typical link, i.e.,
\begin{align} \label{equ:BarDelta_Anlys}
\bar{ \Delta } = \mathbb{E} \left[ \bar{ \Delta }_0 \right].
\end{align}
Note that if every packet can be successfully delivered upon each transmission attempt (namely, when the AoI of a source node reaches threshold $A$), the time average AoI across the network is $(A+1)/2$.
This idle scenario provides a \textit{fundamental lower bound} to the network average AoI, i.e., 
\begin{align} 
\bar{ \Delta } \geq \frac{ A + 1 }{ 2 }.
\end{align}

\begin{figure}[t!]
  \centering{}

    {\includegraphics[width=0.95\columnwidth]{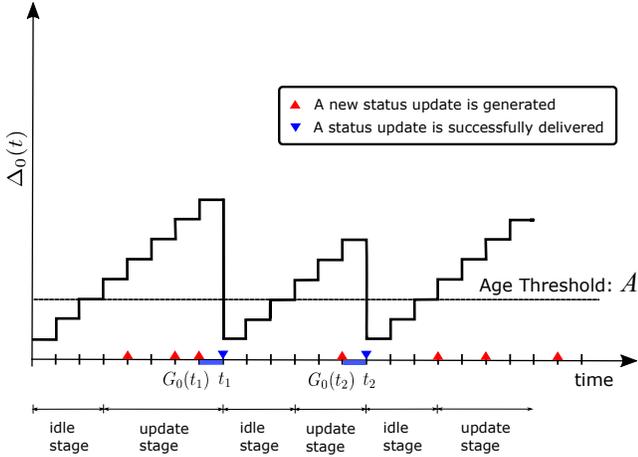}}

  \caption{ An example of the time evolution of age over the link under consideration. The time instance $G_0(t_i)$ and $t_i$ denote the moment when the $i$-th packet is generated and delivered, respectively, and the age is reset to $t_i - G_0(t_i)$. }
  \label{fig:AoIMod_V1}
\end{figure}

\section{Analysis of Average AoI} \label{Sec:AveAoI_Analysis}

\subsection{Preliminaries}
\subsubsection{SINR at the typical receiver}
If the typical source sends out an information packet at time slot $t$, the SINR received at the destination can be written as:
\begin{align} \label{equ:SINR}
\mathrm{SINR}_{0}(t) = \frac{ H_{00}(t) r^{-\alpha} }{ \sum_{ j \neq 0 }  H_{j0} (t) \nu_{j}(t) \Vert X_j \Vert^{-\alpha}  + 1/\rho }
\end{align}
where $\alpha$ is the path loss exponent, $\rho$ is the signal-to-noise ratio (SNR), $H_{ji}(t) \sim \exp(1)$ stands for the channel fading from source $j$ to receiver $i$ at time slot $t$, $\Vert \cdot \Vert$ denotes the Euclidean norm, and $\nu_{j}(t) \in \{ 0, 1 \}$ represents the updating decision of node $j$, which is set to $1$ if node $j$ decides to send a new update to the destination at time slot $t$ and $0$ otherwise.

\subsubsection{Conditional transmission success probability}
Because the network is static, we leverage the notion of the conditional transmission success probability \cite{haenggi2016meta} to quantify the transmission quality of each information packet.
Particularly, given the node positions $\Phi \triangleq \tilde{\Phi} \cup \bar{\Phi}$, the conditional transmission success probability of the typical receiver at a given time slot $t$ is defined as \cite{haenggi2016meta,Hae:21CL-1,Hae:21CL-2}
\begin{align}\label{equ:CndTX_Prob}
\mu^\Phi_{0}(t) = \mathbb{P}\big(\mathrm{SINR}_{0}(t) > \theta \mid \Phi\big),
\end{align}
where $\theta$ is the decoding threshold. 
If we view the dynamics over the typical link from the perspective of queueing theory, the quantity $\mu^\Phi_{0}(t)$ provides complete information about how fast packets are delivered over the wireless channel. 
Therefore, we also refer to it as the service rate when there is no ambiguity. 

Since the system has reached the steady state, when we condition on the network topology $\Phi$, events of successfully transmitting an information packet over each time slot are i.i.d. with probability $\mu^\Phi_i = \mu^\Phi_{i}(t)$, $\forall t \geq 0$. 
Consequently, we drop the time index in the following analysis.

\subsubsection{Conditional average AoI}
Using the above results, we can derive a conditional form of the average AoI as follows. 

\begin{lemma} \label{lma:CndAveAoI}
\textit{Given the point process $\Phi$, the time average AoI over the typical link under TSA is given as:
  \begin{align} \label{equ:Cnd_Ave_AoI}
   \bar{ \Delta }_0 = \frac{ A+1 }{ 2 } + \dfrac{ \dfrac{ A-1 }{2} + \dfrac{ 1 }{ \eta \mu^\Phi_0 } }{ 1 + A \eta \mu^\Phi_0 }.
  \end{align}
}
\end{lemma}
\begin{IEEEproof}
Please see Appendix~\ref{apx:CndAveAoI}.
\end{IEEEproof}

The lemma discloses that the conditional transmission success probability $\mu^\Phi_0$ plays an essential role in characterizing the average AoI.
As such, we detail the steps in deriving the distribution of $\mu^\Phi_0$ next.

\subsection{Distribution of the Service Rate}
We can observe from \eqref{equ:SINR} that the randomness in the $\mathrm{SINR}_0$ stems from three sources: ($i$) the channel fading, ($ii$) the active states of interferers, and ($iii$) the spatial topology of the network.
In what follows, we present the steps to average out the randomness in SINR one by one and finally obtain the distribution of the service rate. 

Firstly, due to interference effects, the source nodes' active states are correlated in time, 
hindering tractable analysis. 
To that end, we adopt the following approximation, which replaces the mutual/local interactions with an average or effective global interaction \cite{EmaElSBau:20,LuSalHae:21}.

\textbf{Approximation~1:}
\textit{ The source nodes experience independent interference over time, and hence their active states are independent of each other. }

The accuracy of this approximation will be verified in Fig.~\ref{fig:VerAnaly}. 
Following Approximation~1, we can obtain an initial expression for the conditional transmission success probability. 

\begin{lemma} \label{lma:Cnd_SucProb}
\textit{Conditioned on the point process $\Phi$, the probability of successful transmission over the typical link is
\begin{align}\label{equ:mu_0Phi}
\mu^\Phi_0 = e^{-\frac{ \theta r^\alpha}{\rho} } \prod_{ i \neq 0 } \Big( 1 - \frac{ a^\Phi_i }{ 1 + \Vert X_i \Vert^\alpha / \theta r^\alpha } \Big)
\end{align}
where $a^\Phi_i = \lim_{ T \rightarrow \infty} \sum_{t=0}^{T-1} \nu_{i,t}/T$ denotes the conditional active probability of node $i$, i.e., the fraction of time that node $i$ is activated and accessing the radio channel.
}
\end{lemma}
\begin{IEEEproof}
The proof is similar to Lemma~3 of \cite{YanXuWan:19} and is omitted here.
\end{IEEEproof}

Next, let us calculate the chance that source node $i$ is interfering with the typical receiver at a given time slot:
\begin{lemma} \label{lma:CndActProb}
\textit{Conditioned on the point process $\Phi$, the active probability of source node $i$ is 
  \begin{align}
  a^\Phi_i = \frac{ \eta }{ 1 + A \eta \mu^\Phi_i }.
  \end{align}
}
\end{lemma}
\begin{IEEEproof}
Please see Appendix~\ref{apx:CndActProb}.
\end{IEEEproof}
And we denote the unconditioned active probability of the typical source node as 
\begin{align}\label{equ:AveActProb_Def}
\varphi &= \mathbb{E}\left[ a^\Phi_0 \right].
\end{align}

Combining the two lemmas, we have the following theorem.
\begin{theorem} \label{thm:CDF_ServRt}
\textit{The cumulative distribution function (CDF) of the service rate satisfies 
\begin{align} \label{equ:Meta_Grl}
F(u) &= \mathbb{P}( \mu^\Phi_0 \geq u )
\nonumber\\
&= \frac{1}{2} -\! \int_{0}^{\infty} \!\!\! \mathrm{Im}\bigg\{ \exp\!\Big(\! - j \omega \log u -  j \omega \frac{\theta r^\alpha }{ \rho } 
\nonumber\\
& -  \lambda \pi r^2 \theta^{ \delta } \Omega_\delta \sum_{k=1}^{\infty} \binom{j \omega}{ k } \binom{ \delta - 1 }{ k - 1 }  \! \int_{0}^1 \!\! \frac{ \eta^k dF(s) }{ ( 1 +\! A \eta s )^k } \Big) \bigg\} \frac{ d \omega }{ \pi \omega }
\end{align}
where $j = \sqrt{-1}$, $\delta = 2/\alpha$, $\mathrm{Im}\{ \cdot \}$ denotes the imaginary part of a complex quantity, and 
\begin{align}
\Omega_\delta = \frac{ \pi \delta }{ \sin \big( \pi \delta \big) }.
\end{align}
}
\end{theorem}
\begin{IEEEproof}
Please see Appendix~\ref{apx:CDF_ServRt}.
\end{IEEEproof}

The CCDF of $\mu^\Phi_0$, $1-F(u)$, is commonly known as the SINR meta distribution \cite{haenggi2016meta,Hae:21CL-1,Hae:21CL-2}. 
It quantifies the fraction of source-destination pairs in the network that attain an SINR that surpasses the threshold $\theta$ with probability at least $u$. 
Due to the space-time interactions among the nodes, the expression of $F(u)$ given in \eqref{equ:Meta_Grl} is a fixed-point functional equation. 

It is worthwhile to write down the first moment of $\mu^\Phi_0$, a.k.a. the transmission success probability, which has been widely used to assess the performance of wireless links.
\begin{corollary}
\textit{When the network is interference limited, i.e., $\frac{1}{\rho} \rightarrow 0$, the transmission success probability $P_s(\theta) = \mathbb{E}[ \mathbb{P}(\mathrm{SIR}_{0} > \theta | \Phi) ]$ can be expressed as 
  \begin{align} \label{equ:CovProb}
  P_\mathrm{s}(\theta) &= \exp\left( - \lambda \pi r^2 \theta^\delta \varphi \Omega_\delta \right) \\ \label{equ:CovProb_FirsApprox}
  & \approx \exp\left( - \frac{ \lambda \pi r^2 \theta^\delta \eta \Omega_\delta }{  1 + A \eta P_\mathrm{s}(\theta) } \right),
  \end{align}
  where $\varphi$ is defined in \eqref{equ:AveActProb_Def} and can be calculated as
  \begin{align} \label{equ:AveActProb}
  \varphi = \int_0^1 \frac{ \eta d F(u) }{ 1 + A \eta u }.
  \end{align}
}
\end{corollary}
\begin{IEEEproof}
By assigning $m=1$ in \eqref{equ:muPhi_m}, we have \eqref{equ:CovProb}; and \eqref{equ:CovProb_FirsApprox} follows by applying Jensen's inequality. 
\end{IEEEproof}

Following this corollary, we consider two extreme operation regimes of the age threshold to garner better insights into the TSA protocol. Particularly, when $A=0$, we have 
\begin{align}
P_\mathrm{s}(\theta) = \exp ( - \lambda \pi r^2 \theta^\delta \eta \Omega_\delta ),
\end{align}
which is the well-known transmission success probability under the slotted ALOHA protocol \cite{HaeAndBac:09}. 
On the other hand, if $A \gg 1$, we have from \eqref{equ:CovProb_FirsApprox} 
\begin{align}
\frac{1}{ P_\mathrm{s}(\theta) } \approx \exp\left( \frac{ \lambda \pi r^2 \theta^\delta \eta \Omega_\delta  }{  1 + A \eta P_\mathrm{s}(\theta) } \right)
\approx 1 + \frac{ \lambda \pi r^2 \theta^\delta \eta \Omega_\delta }{  P_\mathrm{s}(\theta) + A \eta P_\mathrm{s}(\theta) }, 
\end{align}
and solving this fixed-point equation yields 
\begin{align}
P_\mathrm{s}(\theta) \approx 1 - \frac{ \lambda \pi r^2 \theta^\delta \eta \Omega_\delta }{  1 + A \eta }.
\end{align}

\subsection{Average AoI}
Next, we present the analytical expression for the average AoI.
\begin{theorem}
\textit{The network average AoI under the TSA protocol is given by 
  \begin{align} \label{equ:AveAoI}
    \bar{\Delta} = \frac{ A \!+\! 1 }{ 2 } \Big( 1 - \frac{\varphi }{ \eta } \Big) + \frac{\exp\!\Big( \frac{ \theta r^\alpha }{ \rho } \!+\! \frac{ \lambda \pi r^2 \theta^\delta \varphi \Omega_\delta }{ \left( 1-\varphi \right)^{ 1 - \delta } } \Big)}{ \eta }, 
  \end{align}
  where $\varphi(\eta, A)$ is given in \eqref{equ:AveActProb}.
}
\end{theorem}
\begin{IEEEproof}
According to \eqref{equ:BarDelta_Anlys}, calculating $\bar{\Delta}$ is equivalent to computing $\mathbb{E}[ \bar{\Delta}_0 ]$. 
Using the expression in \eqref{equ:Cnd_Ave_AoI}, we can expand $\bar{ \Delta }_0$ as follows: 
\begin{align}
\bar{\Delta}_0 = \frac{ A + 1 }{2} \times \Big( 1 - \frac{ 1 }{ 1 + A \eta \mu^\Phi_0 } \Big) + \frac{ 1 }{ \eta \mu^\Phi_0 }.
\end{align}
Then, by taking an expectation on both sides of the above equation, we have: 
\begin{align}
\mathbb{E}\left[ \bar{\Delta}_0 \right] = \frac{ A + 1 }{2} \times \Big( 1 - \frac{ \varphi }{ \eta } \Big) + \frac{ 1 }{ \eta } \mathbb{E} \Big[\, \frac{ 1 }{ \mu^\Phi_0 } \,\Big], 
\end{align}
and the result follows by taking $m = -1$ in \eqref{equ:muPhi_m}.
\end{IEEEproof}

The expression given in Theorem~2 accounts for the effects from the temporal perspective such as the status updating rate and age threshold, as well as the spatial perspective, i.e., the deployment density, topology, and interference, on the age performance.
We can see from \eqref{equ:AveAoI} that these network parameters jointly influence the average AoI in a composite manner. 
In order to gather more insights, we study a few special cases as follows.  

\subsubsection{Very lazy updating} If the source nodes update to their destinations in a very low frequency, i.e., $\eta \rightarrow 0$, we have 
\begin{align}
\frac{ \varphi }{ \eta } \rightarrow 1
\end{align}
which results in $\bar{\Delta} \rightarrow \infty$, according to \eqref{equ:AveAoI}.
This observation aligns with the current consensus about AoI that a prolonged update rate is detrimental to age performance.

\subsubsection{Very aggressive updating} If the source nodes update to their destinations in a very high frequency, i.e., $\eta \rightarrow 1$, we have 
\begin{align}
\varphi = \int_0^1 \frac{ d F(u) }{ 1 + Au } < 1  
\end{align}
and
\begin{align}
1 - \varphi &= \int_0^1 \frac{ A u }{ 1 + Au } d F(u)  
\nonumber\\
& \geq \int_0^1 \frac{u}{ 1 + u } d F(u) \geq \frac{ P_\mathrm{s}(\theta) }{2}.
\end{align}
Consequently, the network average AoI can be bounded as 
\begin{align}
    \bar{\Delta} < \frac{A}{2} + \exp\!\bigg( \frac{ \theta r^\alpha }{ \rho } \!+\! \frac{ \lambda \pi r^2 \theta^\delta \Omega_\delta  }{ \left( P_\mathrm{s}(\theta) / 2 \right)^{ 1 - \delta } } \bigg).
\end{align}
This result indicates that integrating an age threshold into the slotted ALOHA protocol can effectively alleviate the severe interference in the aggressive updating scenario, which could result in an unbounded network average AoI (we will detail this phenomenon in the subsequent case).

\subsubsection{Sparse deployment} When the nodes are sparsely deployed, i.e., $\lambda \rightarrow 0$, the network is in the noise-limited regime. 
As such, the SINR in \eqref{equ:SINR} reduces to 
\begin{align}
\mathrm{SINR}_0 \approx \rho H_{00} r^{-\alpha}
\end{align}
and the link service rate is given by 
\begin{align}
\mu^\Phi_0 = \mathbb{P}\left( \mathrm{SINR}_0 > \theta \mid \Phi \right) = \exp \Big( - \frac{ \theta r^\alpha }{ \rho } \Big).
\end{align}
As a consequence, we have 
\begin{align} \label{equ:AveAoI_SparseDeply}
\bar{\Delta} = \frac{ A+1 }{ 2 } + \dfrac{ \frac{ A-1 }{2} + \frac{ \exp\big( \frac{ \theta r^\alpha }{ \rho } \big) }{ \eta } }{ 1 + A \eta e^{ - \frac{ \theta r^\alpha }{ \rho } } }.
\end{align}
If we take the derivative of $\bar{\Delta}$ with respect to $A$ in \eqref{equ:AveAoI_SparseDeply}, we obtain
\begin{align}
\frac{ \partial \bar{\Delta} }{ \partial A } = \frac{ A \eta e^{ - \frac{ \theta r^\alpha }{ \rho } }  + A^2 \eta^2 e^{ - \frac{ 2 \theta r^\alpha }{ \rho } }  }{ 2 \big( 1 + A \eta e^{ - \frac{ \theta r^\alpha }{ \rho } } \big)  } > 0,
\end{align}
which implies that the age threshold is not effective in reducing AoI in sparse networks. 
Indeed, if the influence from interference is mild, one shall rev up the updating rate at every source node to promote fresh information delivery.

\subsubsection{No age threshold} If the network operates without an age threshold, i.e., $A=0$, the node activation probability reduces to $\varphi(\eta, 0) = \eta$.
By substituting it into \eqref{equ:AveAoI}, we have: 
\begin{align} \label{equ:BlindAoI}
\bar{\Delta} = \frac{1}{ {\eta} } \exp\Bigg(\, \frac{ \theta r^\alpha }{ \rho } \!+\! \frac{ \lambda \pi r^2 \theta^\delta \eta \Omega_\delta }{ \left( 1-\eta \right)^{ 1 - \delta }} \,\Bigg).
\end{align}
It can be seen from \eqref{equ:BlindAoI} that without an age threshold, aggressively increasing the status updating rate $\eta$ can be deleterious to the AoI, irrespective of the network density (to see this, let $\eta \rightarrow 1$ and it gives $\bar{\Delta} \rightarrow \infty$).
The main reason is ascribed to the positive probability of having an interferer arbitrarily close. 
Therefore, if nodes are updating at a very high frequency, one shall impose an age threshold to reduce the updating frequency.

On the other hand, we can adjust the parameter $\eta$ to optimize AoI in this scenario. 
Specifically, we can take a derivative on the right hand side of \eqref{equ:BlindAoI} with respect to $\eta$, assign it to zero, and solve for the optimal updating rate at each node, which is given by the solution to 
\begin{align}
\lambda \pi r^2 \theta^\delta \Omega_\delta \eta ( 1 - \delta \eta ) ( 1 - \eta )^{\delta - 2} = 1.
\end{align}

\subsubsection{Large age threshold} 
For fixed $\eta$, we have 
\begin{align}
\bar{\Delta} \sim \frac{ A }{ 2 }, \quad A \rightarrow \infty,
\end{align}
which indicates that for large $A$, the network average AoI goes up monotonically with the age threshold. In other words, raising the age threshold too high does not benefit the AoI performance. 
This result also implies that the time average AoI approaches the fundamental lower bound $(A+1)/2$ as $A \rightarrow \infty$.

\section{Simulation and Numerical Results} \label{Sec:SimNum_Results}
In this section, we provide simulation results to validate the accuracy of the approximations in our analytical framework. Based on the analysis, we further investigate the AoI and delay performance under different settings of network parameters.
Particularly, we consider a square region with a side length of $1,000$ unit length, in which source-destination pairs are scattered according to a Poisson bipolar network with spatial density $\lambda$.
The topology remains unchanged once it is generated.
To eliminate the favorable interference coordinations induced by network edges, we use wrapped-around boundaries \cite{FasMueRup:19} that allow dipoles that leave the region on one side to reappear on the opposite side, thus mirroring the missing interferers beyond the scenario boundary.
Then, the dynamics of status updates over each link are run over $10,000$ time slots.
Specifically, every source node records the age information locally and remains silent when the age is below the threshold $A$. 
If the AoI of a source node exceeds threshold $A$, it turns into the updating mode: at the beginning of each time slot, channel gains are independently instantiated, and status updates are generated with probability $\eta$. A packet is successfully received if the SINR at the intended destination exceeds the decoding threshold.
The AoI statistics of the receivers of all the links are recorded to construct the average AoI.
Unless differently specified, we use the following parameters: $r=2.5$, $\lambda=5 \times 10^{-2}$, $\alpha = 3.8$, $\theta=0$~dB, $P_{\mathrm{tx}}=17$~dBm, and $\sigma^2 = -90$~dBm.

\begin{figure}[t!]
  \centering{}

    {\includegraphics[width=0.9\columnwidth]{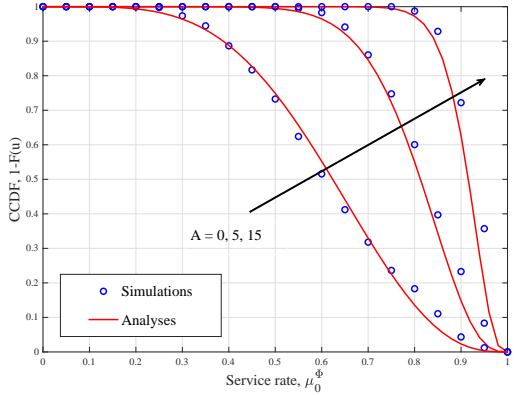}}

  \caption{Simulation versus analysis: the SINR meta distribution, $1-F(u)$. The update rate is set as $\eta=0.3$ and the age threshold varies as $A = {0, 5, 15}$. }
  \label{fig:VerAnaly}
\end{figure}

Fig.~\ref{fig:VerAnaly} compares the simulated CCDF of the conditional transmission success probability to the analysis presented in Theorem~\ref{thm:CDF_ServRt}, for various values of the age threshold. 
We notice the closed match between the analytical results and simulations, which validates Approximation~1 used in our mathematical derivation. 
This figure also provides fine-grained information about the link qualities across the network. 
For instance, in the depicted situation, when $A=0$ (in this case, TSA reduces to the conventional slotted ALOHA protocol) only $3\%$ of the link pairs can achieve the desired SINR, while that portion sheers up to $70\%$ when $A$ increases to $15$. 
Therefore, adopting an age threshold into the slotted ALOHA is effective in boosting up the link performance. 

\begin{figure}[t!]
  \centering{}

    {\includegraphics[width=0.9\columnwidth]{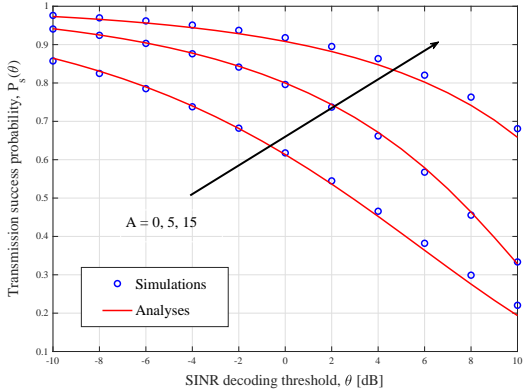}}

  \caption{Transmission success probability versus SINR decoding threshold. The update rate is set as $\eta=0.3$, and the age threshold varies as $A = {0, 5, 15}$. }
  \label{fig:VerAnaly_CovProb}
\end{figure}

In Fig.~\ref{fig:VerAnaly_CovProb}, we plot the typical node's transmission success probability as a function of the SINR decoding threshold under a variety of age thresholds.
We can see that incorporating an age threshold into the slotted ALOHA protocol is beneficial for enhancing link performance. 
This can be ascribed to the effectiveness of the age threshold in mitigating the interference.

\begin{figure}[t!]
  \centering{}

    {\includegraphics[width=0.9\columnwidth]{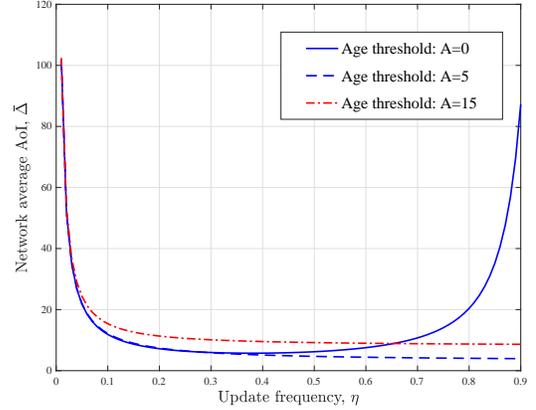}}

  \caption{Network average AoI under different age thresholds.}
  \label{fig:AveAoI_UpdtRt}
\end{figure}

Fig.~\ref{fig:AveAoI_UpdtRt} depicts the network average AoI as a function of the status updating rate $\eta$, for a set of ascending values of age threshold. 
We can see that when the network operates under slotted ALOHA without an age threshold, an optimal frequency for status updating exists that minimizes the average AoI of the typical source node. 
This is achieved by striking a balance between ($a$) information freshness at the source and ($b$) spatial contention
in the radio access channels across the network. 
On the other hand, by imposing an age threshold at the transmitters, the receivers enjoy a leveled-down interference and hence can rev up their update rates to achieve a smaller AoI.
Nonetheless, by comparing the cases between $A=5$ and $A=15$, we notice that merely increasing the age threshold is not always beneficial for reducing the average AoI. Therefore, the age threshold shall be carefully optimized. 

\begin{figure}[t!]
  \centering{}

    {\includegraphics[width=0.85\columnwidth]{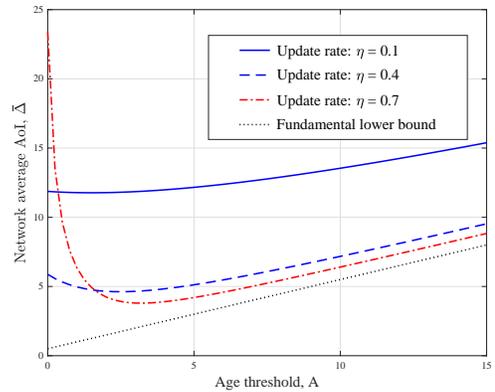}}

  \caption{Network average AoI versus age threshold, under different frequencies of status updating.}
  \label{fig:AveAoI_AgeThreshold}
\end{figure}

As a consequence, we draw the network average AoI as a function of the age threshold in Fig.~\ref{fig:AveAoI_AgeThreshold}.
The results of this figure demonstrate that for a given status update rate, an optimal age threshold exists that minimizes the network average AoI.
Moreover, by contrasting the situations under rare updating, i.e., $\eta = 0.1$, and frequent updating, i.e., $\eta = 0.7$, we note that employing the TSA protocol at the source nodes and correspondingly increasing their update rates can efficaciously reduce the average AoI. 
These observations corroborate the effectiveness of the slotted ALOHA with an age threshold in reducing the average AoI.

\begin{figure}[t!]
  \centering{}

    {\includegraphics[width=0.9\columnwidth]{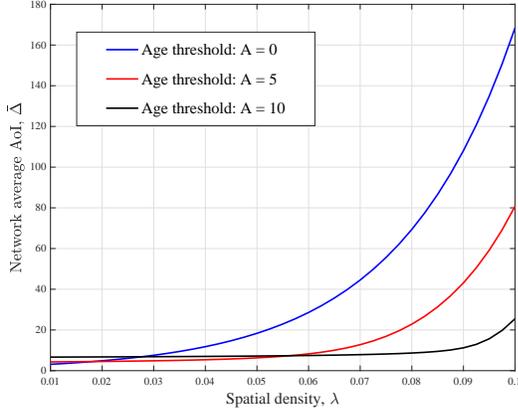}}

  \caption{Average and variance of AoI over the network versus spatial density. The update rate is set as $\eta = 0.5$ and the transceiver distance is $r=3.5$.}
  \label{fig:AoI_vs_SptlDensity}
\end{figure}

To investigate the joint effect of network densification and TSA protocol on the age performance, we plot the average, as well as the variance of AoI for an ascending value of deployment density in Fig.~\ref{fig:AoI_vs_SptlDensity}. 
From Fig.~\ref{fig:AoI_vs_SptlDensity}(a), we observe that despite densifying the infrastructure inevitably increases the network average AoI in all the considered cases, the one under the conventional slotted ALOHA surge rapidly while those under the TSA protocol goes up steadily. 
Particularly, with a ten-fold increase in the deployment density, TSA protocol with $A=10$ attains a network average AoI smaller than that under the slotted ALOHA by order of magnitude. 
The reason for this is that imposing an age threshold at the source nodes can effectively alleviate the spatial contention among the transmitters, facilitating information delivery in the network by controlling interference.

\section{Conclusion} \label{Sec:Conclusion}
In this paper, we conducted an analytical study toward understanding the effect of TSA protocol on the age performance of a large random access network. 
We established a theoretical framework that accounts for update rate, age threshold, signal propagation factors, and interference. 
We derived analytical expressions for the SINR meta distribution as well as the average AoI. 
Based on the analysis, we thoroughly explored the effect of TSA on network performance. 
Specifically, we confirmed that incorporating an age threshold into the slotted ALOHA protocol is beneficial for reducing the average AoI, whereas the gain is particularly pronounced when the network is densely deployed. 
Nonetheless, one needs to jointly optimize the update rate and age threshold to fully exploit the potential of TSA.

\begin{appendix}
\subsection{Proof of Lemma~\ref{lma:CndAveAoI} } \label{apx:CndAveAoI}
Let us denote $I_n$ as the time interval between two consecutive transmission attempts over the typical link and $J_k$ the waiting time at the destination between successful receptions of the $k$-th and ($k+1$)-th updates, respectively. 
We have the following relationship between $I_n$ and $J_k$ according to the age threshold-based transmission protocol:
\begin{align}
  J_k = A + \sum_{ n=1 }^N I_n
\end{align}
where $N$ is a random variable that represents the number of attempts between two successful transmissions. 
We further introduce a variable $L_k$, which represents the area under the AoI evolution curve across the $k$-th successful update, as follows:
\begin{align}
L_k = \sum_{ j = 1 }^{J_k} j = \frac{1}{2} J_k (J_k + 1).
\end{align}

Then, given the point process $\Phi$, we can calculate the conditional average AoI over the typical link as 
\begin{align} \label{equ:Intmd_CndAveAoI}
\bar{\Delta}_0 &= \frac{ \mathbb{E} \big[ L_k | \Phi \big] }{ \mathbb{E} \big[ J_k | \Phi \big] } 
\nonumber\\
&= \frac{1}{2} + \frac{ \mathbb{E} \big[ J^2_k | \Phi \big] }{ 2 \mathbb{E} \big[ J_k | \Phi \big] }.
\end{align}

Moreover, notice that $N$ follows a geometric distribution with parameter $\mu^\Phi_0$, we can compute $\mathbb{E}[ J_k | \Phi ]$ and $\mathbb{E}[ J^2_k | \Phi ]$ respectively as:
\begin{align} \label{equ:EJ_k}
\mathbb{E}\big[ J_k | \Phi \big] &= A + \sum_{ n = 1 }^\infty \mathbb{E}\big[ I_n \big] n \mu^\Phi_0 \big( 1 - \mu^\Phi_0 \big)^{n-1} 
\nonumber\\
&= A + \frac{ \mathbb{E}\big[ I_n \big] }{ \mu^\Phi_0 }
\end{align}
and 
\begin{align} \label{equ:EJ2_k}
\mathbb{E} \big[ J^2_k | \Phi \big] &= A^2 \!+\! 2 A \mathbb{E} \Big[ \sum_{n=1}^N I_n \big| \Phi \Big] \!+\! \mathbb{E} \Big[ \Big(  \sum_{n=1}^N I_n \Big)^2 \big| \Phi \Big]
\nonumber\\
&= A^2 \!+\! \frac{ 2 A \mathbb{E}[ I_n ] + E[ I^2_n ] }{ \mu^\Phi_0 } \!+\! \frac{ 2 \big( 1 - \mu^\Phi_0 \big) \big( \mathbb{E}[I_n] \big)^2 }{ \big( \mu^\Phi_0 \big)^2 }.
\end{align}

According to the TSA protocol, when a source node is allowed to transmit, it generates new updates with probability $\eta$ independently over time; thus, we have
\begin{align} \label{equ:EIn}
\mathbb{E}[ I_n ] &= \frac{ 1 }{ \eta }, \\ \label{equ:EI2n}
\mathbb{E}[ I^2_n ] &= \frac{ 2 - \eta }{ \eta^2 }.
\end{align}
The proof is completed by substituting \eqref{equ:EJ_k}, \eqref{equ:EJ2_k}, \eqref{equ:EIn}, \eqref{equ:EI2n} into \eqref{equ:Intmd_CndAveAoI}.

\subsection{Proof of Lemma~\ref{lma:CndActProb} } \label{apx:CndActProb}
Without loss of generality, let us consider the time interval between the $k$-th and ($k+1$)-th successful updates. 
If we position a time slot $t$ uniformly at random within this interval, the probability that the AoI $\Delta_i(t)$ exceeds threshold $A$ is given by 
\begin{align}
\mathbb{P}\left( \Delta_i(t) > A \right) = \frac{ \frac{ 1 }{ \eta \mu^\Phi_i } }{ A + \frac{ 1 }{ \eta \mu^\Phi_i } }.
\end{align}
As the event that node $i$ being activated at time slot $t$ needs to satisfy two conditions: ($a$) the AoI of node $i$ exceeds threshold $A$ and ($b$) the node generates a new update; we have 
\begin{align}
a^\Phi_i = \eta \, \mathbb{P}\left( \Delta_i(t) > A \right).
\end{align} 
The result follows from further simplifying the above.

\subsection{Proof of Theorem~\ref{thm:CDF_ServRt} } \label{apx:CDF_ServRt}
To obtain the distribution of $\mu^\Phi_0$, we start by calculating its $m$-th moment, given by the following: 
\begin{align} \label{equ:MY_moment}
&\mathbb{E}\big[ ( \mu^\Phi_0 )^m \big]
\nonumber\\
& = e^{ - \frac{m \theta r^\alpha}{ \rho } } \mathbb{E}\Big[ \prod_{ i \neq 0 } \! \big( 1 - \frac{ 1 }{ 1 \!+\! \Vert X_i \Vert^\alpha / \theta r^\alpha } \cdot \frac{ \eta }{ 1 \!+\! A \eta \mu^\Phi_i } \big)^m \Big]
\nonumber\\
& \stackrel{(a)}{=} e^{ - \frac{m \theta r^\alpha}{ \rho } } e^{- \lambda \int_{ \boldsymbol{x} \in \mathbb{R}^2 } \mathbb{E} \Big[ 1 - \big( 1 - \frac{1}{ 1 + \Vert \boldsymbol{x} \Vert^\alpha / \theta r^\alpha } \cdot \frac{ \eta }{ 1 + A \eta \mu_{\boldsymbol{x}} } \big)^m \Big] d\boldsymbol{x} }
\nonumber\\
& \stackrel{(b)}{=} \exp \! \bigg(\! - \frac{ m \theta r^\alpha }{ \rho } - \lambda \! \int_{ \boldsymbol{x} \in \mathbb{R}^2 } \sum_{ k = 1 }^{m} \! \binom{ m }{ k }   \frac{ (-1)^{k+1} d \boldsymbol{x} }{ ( 1 \!+\! \Vert \boldsymbol{x} \Vert^\alpha \!/ \theta r^\alpha )^k }
\nonumber\\
&\qquad \qquad \qquad \qquad \qquad \qquad \times \underbrace{\mathbb{E} \Big[ \big( \frac{ \eta }{ 1 + A \eta \mu_{\boldsymbol{x}} } \big)^k \Big] }_{Q_1} \bigg),
\end{align}
where ($a$) follows by using the probability generating functional (PGFL) of PPP and ($b$) by expanding the expression via binomial theorem.

Due to ergodicity, the conditional transmission success probability $\{ \mu^\Phi_i \}_{ i \in \mathbb{N} }$ are i.i.d. across the transmitters.
As such, let us assume the distribution of $\mu^\Phi_0$, denoted as $F(\cdot)$, is available. 
Then, $Q_1$ can be computed as: 
\begin{align} \label{equ:Q1}
Q_1 = \int_0^1 \frac{ \eta^k F(ds) }{ ( 1 + A \eta s )^k }.
\end{align}
By substituting \eqref{equ:Q1} into \eqref{equ:MY_moment}, it yields:
\begin{align} \label{equ:muPhi_m}
&\mathbb{E}\big[ ( \mu^\Phi_0 )^m \big] 
\nonumber\\
&= \exp\!\Bigg(\!\! - \frac{ m \theta r^\alpha }{ \rho } - \lambda \pi r^2 \theta^\delta \sum_{ k = 1 }^{m} \! \binom{ m }{ k } \frac{ (-1)^{k+1} \! \int_{0}^{\infty} \! dv }{ ( 1 \!+\! v^{ \frac{\alpha}{2} } )^k }
\nonumber\\
&\qquad \qquad \qquad \qquad \qquad \qquad \qquad\quad~\, \times \int_{0}^{1} \frac{ \eta^k F(ds) }{ \big( 1 + A \eta s \big)^k }  \Bigg)
\nonumber\\
&= \exp\!\Bigg(\!\! - \frac{ m \theta r^\alpha }{ \rho } - \lambda \pi r^2 \theta^\delta \sum_{ k = 1 }^{m} \! \binom{ m }{ k } \binom{ \delta - 1 }{ k - 1 } \frac{ \pi \delta }{ \sin( \pi \delta ) }
\nonumber\\
&\qquad \qquad \qquad \qquad \qquad \qquad \qquad\quad~\, \times \int_{0}^{1} \frac{ \eta^k F(ds) }{ \big( 1 + A \eta s \big)^k }  \Bigg).
\end{align}
Finally, by using the Gil-Pelaez theorem \cite{Gil}, we can derive the CDF of $\mu^\Phi_0$ as:
\begin{align}
F(u) &= \mathbb{P}( \mu^\Phi_0 < u ) 
\nonumber\\
&= \frac{1}{2} - \frac{1}{\pi} \int_{0}^{\infty} \mathrm{Im} \Big\{ u^{ - j \omega } \mathbb{E}\big[ ( \mu^\Phi_0 )^{j \omega} \big] \Big\} \frac{d \omega}{ \omega }.
\end{align}
The statement readily follows by substituting \eqref{equ:muPhi_m} into the above equation.

\end{appendix}

\bibliographystyle{IEEEtran}
\bibliography{bib/StringDefinitions,bib/IEEEabrv,bib/ATB_SALOHA}

\end{document}